# Low-dimensional Flow Models from high-dimensional Flow data with Machine Learning and First Principles

by Nan Deng (IMSIA, ENSTA Paris, IP Paris & LIMSI, UPSaclay), Luc R. Pastur (IMSIA, ENSTA Paris, IP Paris) and Bernd R. Noack (Harbin Institute of Technology)

**Reduced-order modelling and system identification can help us figure out the elementary degrees of freedom and the underlying mechanisms from the high-dimensional and nonlinear dynamics of fluid flow. Machine learning has brought new opportunities to these two processes and is revolutionising traditional methods. We show a framework to obtain a sparse human-interpretable model from complex high-dimensional data using machine learning and first principles.**

The complexity of fluid flow dynamics comes from high-dimensional and nonlinear dynamics. However, in many cases, the complex dynamics can be described by some elementary structures, such as vortical structures in the wake, impinging flows, etc. These typical structures, featuring typical spatial and temporal scales, reveal the underlying mechanisms hidden by the high-dimensionality. Benefited from the powerful feature extraction ability of machine learning, dimensionality reduction has become much easier, and numerous methods have been developed. Meanwhile, nonlinear system identification has become more flexible and intelligent. The dynamics can be not only derived from traditional approaches (stability analysis, Galerkin projection, etc.), but also from pure data-driven regression methods. A combination of the data-driven system identification with the constraints from the traditional method leads to a physics-based reduced-order model (ROM), which provides us with a better understanding of the complex flow and contributes to the design of effective control and optimization.

A semi-supervised modelling methodology, combining an unsupervised pattern recognition using Proper Orthogonal Decomposition (POD) with a supervised system identification under physics-based constraints, is proposed to obtain a least-order mean-field model [1], which can well explain the hidden relation between the fluctuation and the mean flow fields. To be mentioned, based on the classic Galerkin framework, the classic Galerkin framework, a critical optimization relies on considering the symmetry of the mean flow field and the anti-symmetry of the fluctuation flow field, which dramatically simplifies the difficulty of system identification and improves their interpretability.

For instance, this framework is used to a transient flow of the toy system "fluidic pinball" [L1], exhibiting two successive Hopf and pitchfork bifurcations, as shown in the 3D phase portrait of Figure 1. At Re=80, investigated with Direct Numerical Simulation (DNS), starting close to the symmetric steady solution, the flow state first reaches an unstable limit cycle, associated with a statistically symmetric vortex shedding, then gradually loses the statistical symmetry and eventually reaches one of the two stable limit cycles, associated with a statistically asymmetric vortex shedding.

The reduced-order modelling strategy is illustrated in the lower part of Figure 1. The first step is to determine the least-dimensional manifold considering all the transient dynamics. Following a Galerkin approach, the necessary number of degrees of freedom to describe the Hopf bifurcation is three, while it is two for the pitchfork bifurcation. From mean-field considerations, the instability is triggered by anti-symmetric eigenmodes with respect to the x-axis in both the Hopf and the pitchfork bifurcations. By contrast, the distortion of the base flow from the steady solution to the mean-field, represented by the so-called "the shift mode", is symmetric and slaved to the corresponding anti-symmetric active modes.

The model decomposition has a clear purpose of flow feature extraction. With snapshots of the velocity field obtained from the DNS, this process can be purely data-driven. The anti-symmetric active modes for the Hopf and pitchfork bifurcation are extracted from the permanent asymptotic regime by the two leading POD modes and the difference between the two mirror-conjugated asymmetric steady solutions, respectively. The two slaved shift modes are recognized from the difference between the symmetric steady solution and their mean flow fields. As a combination of the two bifurcations, the Galerkin expansion consists of at least five independent, orthogonal modes, whose dynamics are coupled together when the Reynolds number is far beyond the critical value of the pitchfork bifurcation.

Nonlinear system identification for a five-dimensional system is very challenging. 25 linear terms ($l_{ij}$) and 75 quadratic terms ($q_{ijk}$) need to identify. Based on symmetry considerations, more than half of the terms vanish. The key terms (growth rates, frequency, slaving relations, and nonlinearity parameters) for each bifurcation are identified from the linear stability analysis of the steady solution and from the typical time and amplitude scales of the asymptotic dynamics on the limit cycles. The simplest dynamical system is shown in the bottom-left of Figure 1. The remaining cross-terms are identified with a supervised method, using a sparse regression algorithm under constraints (SINDy) [2].

A sparse, easily interpretable, five-dimensional Galerkin model has been derived from an infinite flow system. The main features of the manifold on which the dynamics takes place are correctly identified by this least-order mean-field model. For the interested reader, further details can be found in Deng et al. [1]. Comparing to numerous kinds of ROMs [3], the mean-field model emphasizes the nonlinear dynamics of the base-flow distortion from the fluctuation, which provides a theoretical basis from the Reynolds equation for the ROM. The linear-quadratic dynamics of the corresponding Galerkin system is fully consistent with the quadratic nonlinearities of the Navier-Stokes equations.

This framework can be generalized to other flow configurations or even more complex dynamical regimes, like the quasi-periodic regime of the fluidic pinball at higher Reynolds numbers. Based on the least-dimensional ROM, some additional degrees of freedom could be included, e.g. higher harmonic modes, to allow the energy to flow to smaller scales in the model. It can also be used for nonlinear model-based control by including additional actuating modes. Thanks to the clustering and classification abilities of machine learning, the procedure can be further automatized with automated learning of the state space.

**Links:**
[L1]: http://berndnoack.com/FlowControl.php

**Please contact:**
**Nan Deng**
Nan Deng, IMSIA, ENSTA Paris, IP Paris & LIMSI, UPSaclay, France
nan.deng@ensta-paris.fr

# Figures:

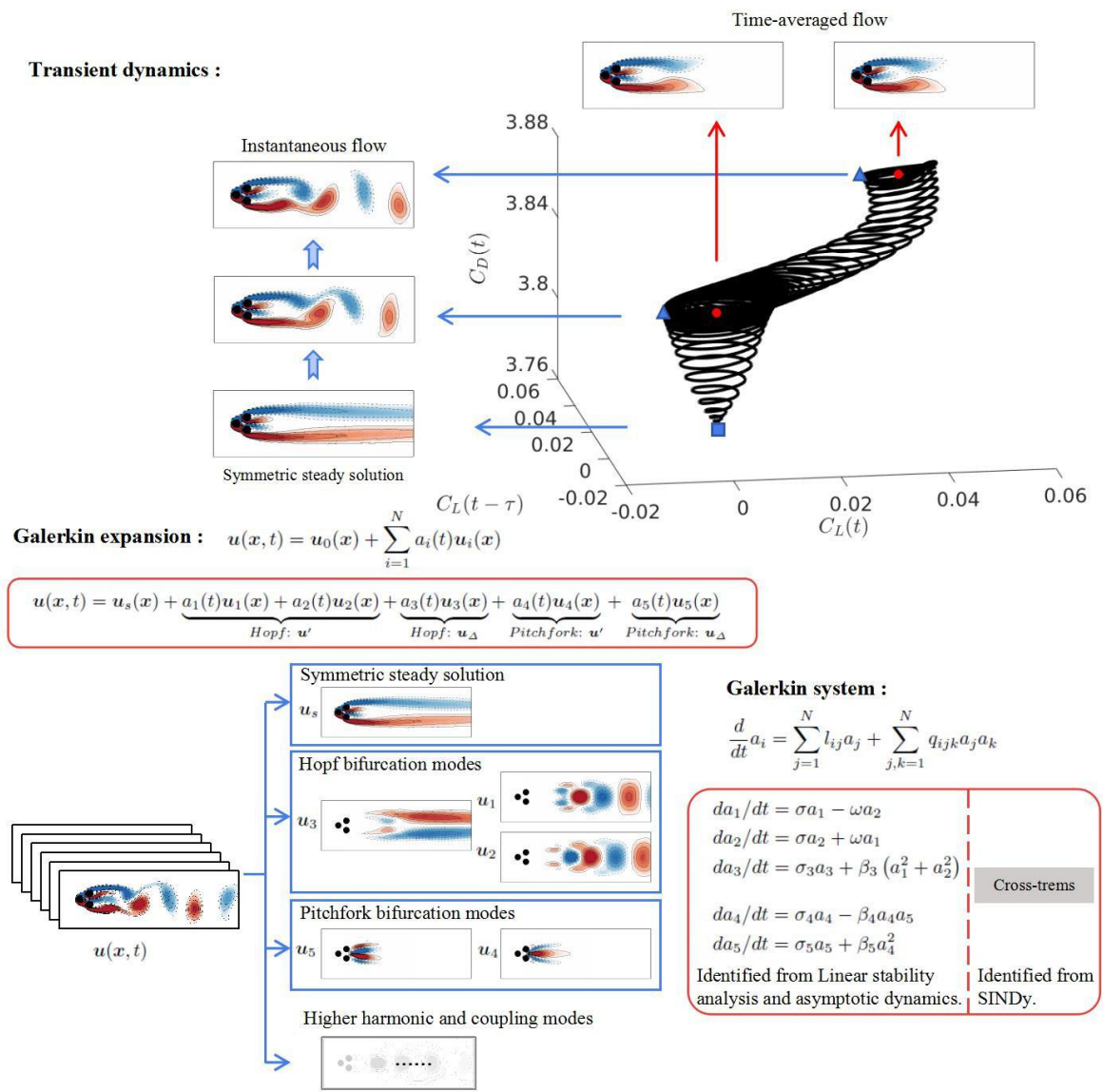

Figure 1. Reduced-order modelling strategy for the fluidic Pinball at Re=80. Upper part: Phase portrait based on the lift coefficient $C_L(t)$, lift coefficient with delay $C_L(t-\tau)$, and drag coefficient $C_D(t)$. A transient scenario starting close to the symmetric steady solution (blue square) first reaches a symmetry-centered limit cycle and then loses the statistical symmetry and approaches to an asymmetry-centered limit cycle. For these two limit cycles, the mean flow field is marked with a red point, and a sampled instantaneous flow field is marked with a blue triangle. Lower part: Mean-field Galerkin expansion and corresponding Galerkin system. In the red boxes is the resulting least order model. The modal decomposition is carried out for the two consecutive bifurcations. Both physics-based and data-driven methods are used during the system identification: the key coefficients in the simplest system are identified from the linear stability

**analysis and the asymptotic dynamics, and the cross-terms are identified from a sparse regression algorithm (SINDy).**